\documentclass[%
 aip,
 amsmath,amssymb,
 reprint,%
]{revtex4-1}

\usepackage{graphicx}
\usepackage{dcolumn}
\usepackage{bm}

\usepackage[utf8]{inputenc}
\usepackage[T1]{fontenc}
\usepackage{mathptmx}
\usepackage{etoolbox}
\usepackage{color}

\newcommand{\COR}[1]{{\color{black} #1}}

\makeatletter
\def\@email#1#2{%
 \endgroup
 \patchcmd{\titleblock@produce}
  {\frontmatter@RRAPformat}
  {\frontmatter@RRAPformat{\produce@RRAP{*#1\href{mailto:#2}{#2}}}\frontmatter@RRAPformat}
  {}{}
}%
\makeatother
\begin{document}

\preprint{AIP/123-QED}

\title[Reconfigurable phase contrast microscopy]{Reconfigurable phase contrast microscopy with correlated photon pairs}
\author{Hazel Hodgson}
 \affiliation{National Research Council of Canada, 100 Sussex Drive, Ottawa, Ontario K1A 0R6, Canada}
\author{Yingwen Zhang}%
\affiliation{Nexus for Quantum Technologies, University of Ottawa, Ontario, K1N 6N5, Ottawa}
\affiliation{National Research Council of Canada, 100 Sussex Drive, Ottawa, Ontario K1A 0R6, Canada}

\author{Duncan England}
 \affiliation{National Research Council of Canada, 100 Sussex Drive, Ottawa, Ontario K1A 0R6, Canada}
 \email{Duncan.England@nrc-cnrc.gc.ca}
 
\author{Benjamin Sussman}
 \affiliation{National Research Council of Canada, 100 Sussex Drive, Ottawa, Ontario K1A 0R6, Canada}
\affiliation{Department of Physics, University of Ottawa, Ottawa, Ontario, K1N 6N5, Canada}

\begin{abstract}
A phase-sensitive microscopy technique is proposed and demonstrated that employs the momentum correlations inherent in spontaneous parametric down-conversion. One photon from a correlated pair is focused onto a microscopic target while the other is measured in the Fourier plane. This provides knowledge of the position and angle of illumination for every photon striking the target, allowing full post-production control of the illumination angle used to form an image. The versatility of this approach is showcased with asymmetric illumination and differential phase contrast imaging, without any beam blocks or moving parts.
\end{abstract}

\maketitle
Optical microscopy was first demonstrated over 400 years ago~\cite{Smith2014}, and the field has been developing and evolving ever since. The majority of these advancements have taken place in the realm of classical optics, where the incident and scattered light are treated as electromagnetic waves. However, recent years have seen the emergence of a series of techniques in which quantum properties of light such as entanglement~\cite{Defienne2021}, or squeezing~\cite{Casacio2021}, can be used to enhance imaging techniques. \COR{When two photons from an entangled pair interact with a sample, image resolution can be improved beyond the diffraction limit~\cite{dangelo_two-photon_2001,israel_supersensitive_2014}. Alternatively, one photon from the pair interacts with the sample, while the other is measured on a separate detector to gain extra information. In this configuration, the resolution cannot be enhanced~\cite{dangelo_resolution_2005}, but new capabilities are offered including ghost imaging~\cite{Padgett2017}, noise supression~\cite{gregory_imaging_2020}, sub-shot-noise imaging~\cite{Brida2010}, interaction-free imaging~\cite{Lemos2014}, hyperspectral imaging~\cite{zhang_snapshot_2022}, and many more~\cite{Genovese2016}.}

In this work, we employ the photon momentum correlations inherent in spontaneous parametric down-conversion (SPDC) to perform phase-sensitive imaging on the micron scale, interrogating both a resolution target and a biological sample. In our setup, photon pairs are produced by SPDC and one photon from each pair is imaged onto the sample and then relay-imaged onto a time-tagging camera to measure the position at which it was generated in the crystal. The second photon from the pair is measured on the same camera in the Fourier plane (far-field) to determine the angle at which it was generated. Photon pairs are identified through temporal coincidences and, through momentum anti-correlation, we can determine the angle at which each photon arrived at the target as well as the position~\cite{Zhang2022}. By applying spatial filtering in the Fourier plane digitally in post-processing, we can achieve full reconfigurability over the range of illumination angles that are used to generate an image. In previous work~\cite{Zhang2022} we used this information to reconstruct a 3-dimensional scene. Here we show that the same principle can be used to make images that are sensitive to the phase gradient of the sample, which modify the photon momentum and can therefore be measured through correlations. 

Our setup is analogous to classical computational imaging techniques in which an active element is used to dynamically control the illumination angle of the scene. Active elements used can include spatial light modulators~\cite{Maurer2011}, or LED arrays~\cite{Liu2014,Tian2014,Tian2015,Lee2015,Jung2017}. Active illumination has been shown to be advantageous in various schemes including dark field imaging~\cite{Goedhart2007}, Fourier Ptycography for digital refocusing and depth sectioning~\cite{Zheng2011,Horstmeyer2016}, motion deblurring~\cite{Ma2015}, and many more~\cite{Maurer2011}. In our work, however, there are no active elements or moving parts, instead the angle of each photon varies randomly due to the stochastic nature of SPDC, and this angle is inferred through momentum anti-correlation measured on the entangled partner photon.   

Of particular relevance to this work is a scheme known as {\em asymmetric imaging} in which a partial beam-block placed before the condenser lens leads to a phase-like image. The partial block can be made by physically blocking half of the beam~\cite{Mehta2009}, or can be digitally generated using the active element~\cite{Tian2014}. In either case, phase gradients in the target lead to bright or dark highlights in the image, thereby enhancing the phase contrast. A related technique known as {\em differential phase contrast (DPC) imaging} subtracts two asymmetric images obtained using complementary illumination patterns to remove the background and further improve the phase contrast~\cite{Tian2015}. In this work we use photon correlation measurements to demonstrate both asymmetric and DPC imaging using arbitrary illumination patterns. Because the angle of each illuminating photon is known, this can be achieved entirely in post-processing after the image has been acquired.

\begin{figure*} [ht]
 \includegraphics[width=0.7\linewidth]{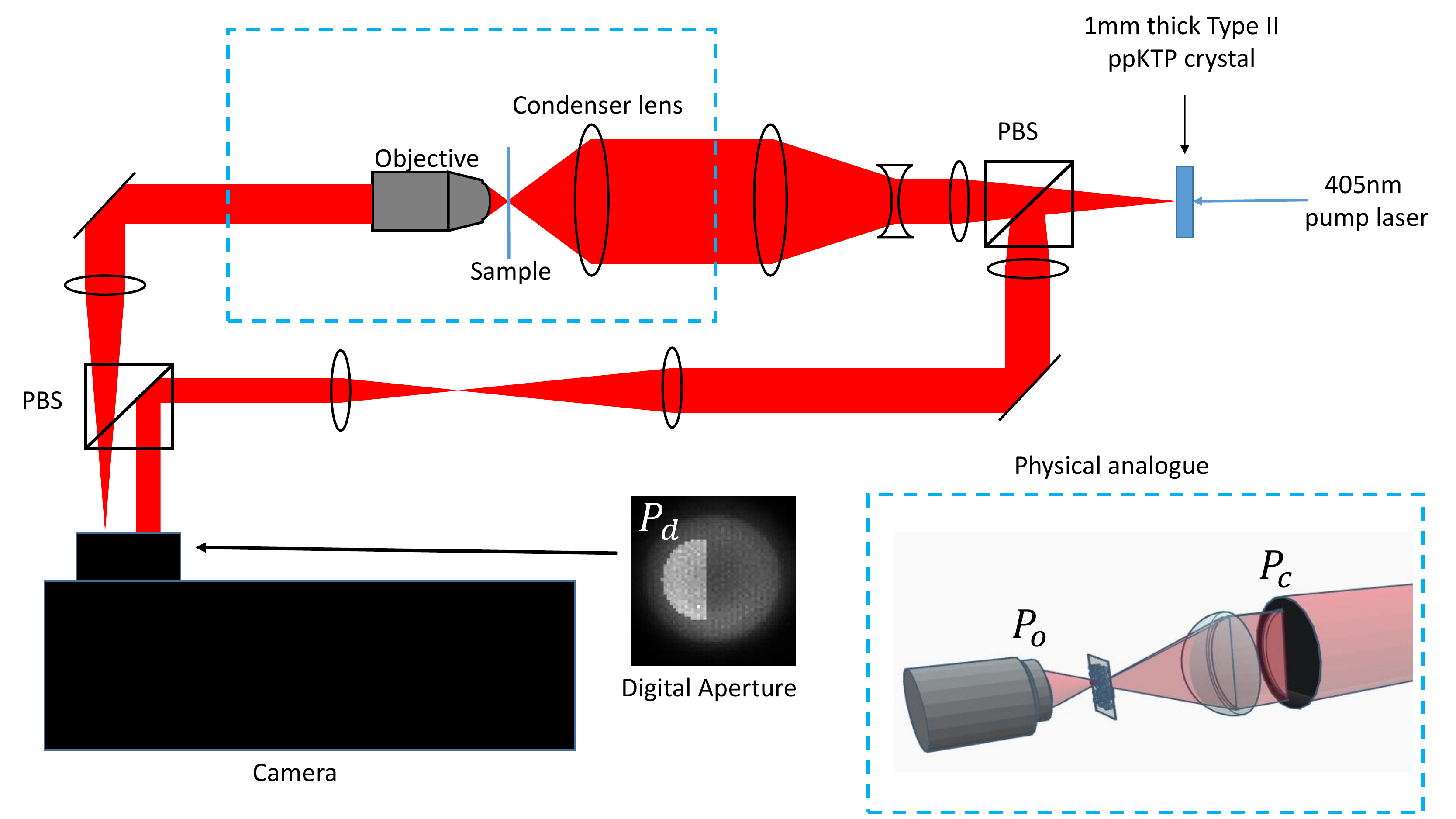}
 \centering
 \caption{Experimental setup. Spatial-temporal correlated photon pairs with orthogonal polarization at 810\,nm are generated through SPDC by pumping a 1\,mm thick type II ppKTP crystal with a 405\,nm CW laser. With the 405\,nm light filtered out using a longpass filter (not shown), the 810\,nm photon pairs are separated into two paths using a PBS. In one path, the image plane of the crystal is first imaged onto the sample and then the camera. In the other path, the Fourier plane of the crystal is directly imaged onto a different part of the same camera. Post-selections in the Fourier plane are used to generate digital apertures. The physical analogue of our post-processing digital selection is shown in the highlighted blue box.
 }
  \label{fig:Setup}
\end{figure*}

\begin{figure*}[ht]
\centering
\includegraphics[width=0.8\linewidth]{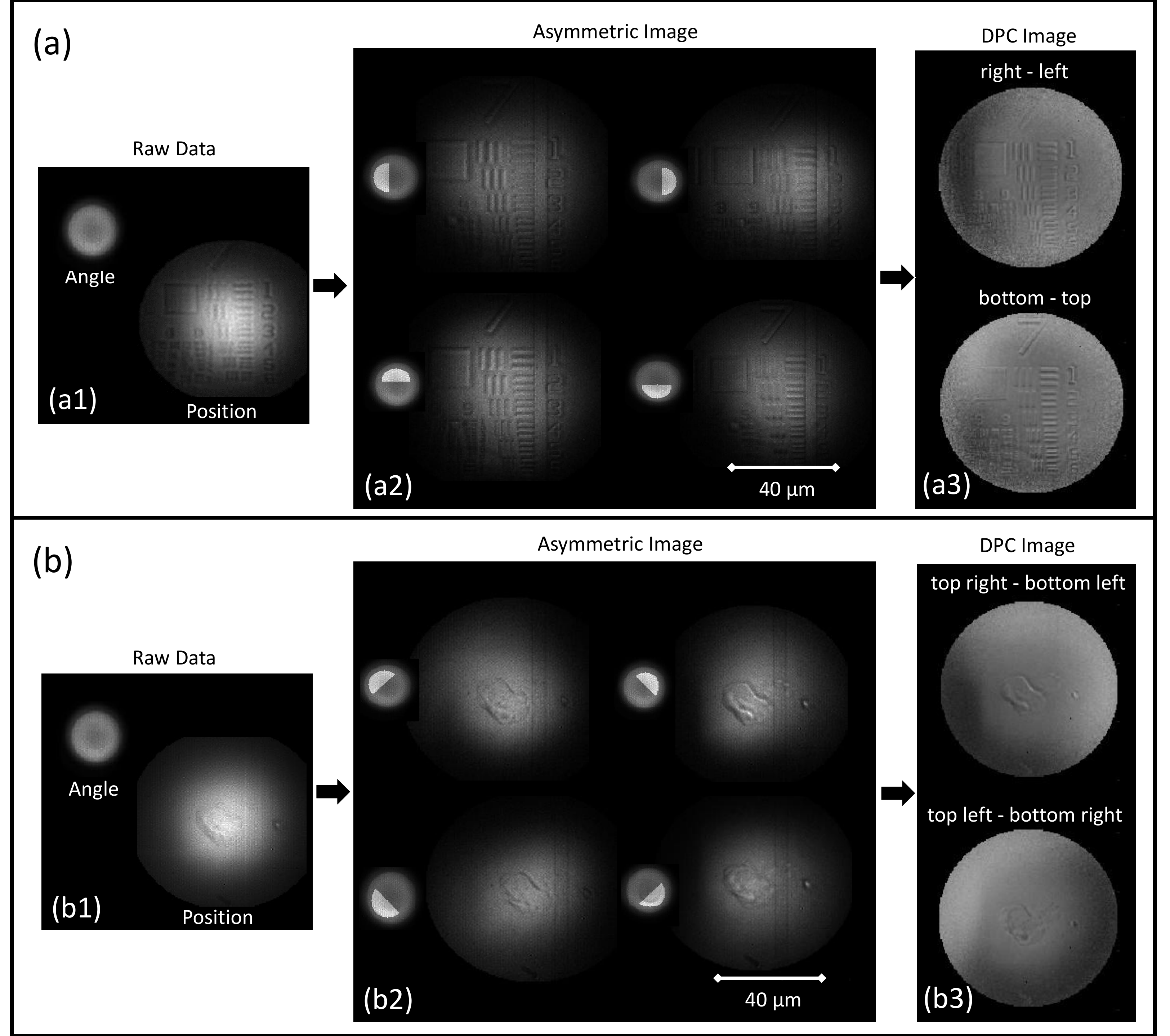}
 \centering
 \caption{Raw data to asymmetric and DPC image transformation. (a) Images of a 1951 USAF resolution phase target with 350\,nm etching depth and (b) images of an unstained cheek cell. (a1) and (b1) are the raw image data showing the idler photon image taken in the crystal far field (angle) and the signal photon image taken in the crystal near field (position). (a2) and (b2) are coincidence images obtained through various asymmetric aperture selections across the diagonal axes of the far field image. (a3) and (b3) are the resultant DPC images.}
\label{fig2}
\end{figure*}

\begin{figure}[ht]
\centering
\includegraphics[width=\linewidth]{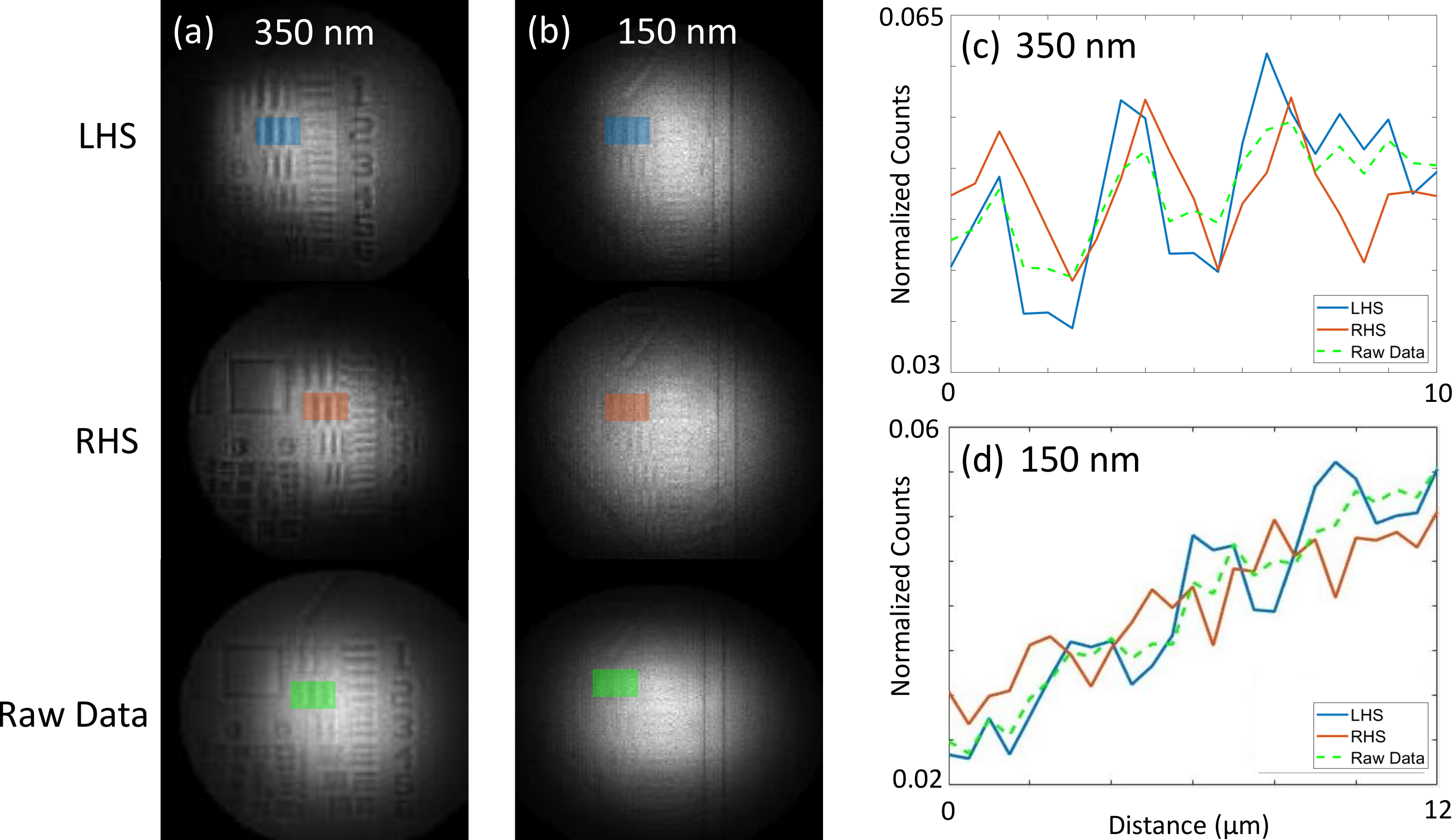}
 \centering
 \caption{Contrast comparison between direct and asymmetric imaging.  Asymmetric imaging taken with left-hemisphere selection (LHS), right-hemisphere selection (RHS) and raw direct imaging for a resolution target with etching depth of (a) 350\,nm and (b) 150\,nm. The region along which the cross-section contrast is compared is highlighted in color with the corresponding cross-section shown in (c) and (d).}
\label{fig:Contrast}
\end{figure}

\COR{The mathematical formalism behind asymmetric imaging is discussed in detail elsewhere~\cite{Hamilton1984,Yi2006,Tian2014,Tian2015}, so only a brief conceptual introduction is given here. The classic setup for asymmetric imaging is shown in the inset of figure~\ref{fig:Setup} where an aperture $P_c$ is placed before the condenser lens and an aperture $P_o$ is placed after the objective lens~\cite{Hamilton1984}. These apertures can be actual beam blocks, as shown in the case of $P_c$, or can simply be defined by the physical size of the optic, as in $P_o$. A sample is placed in the focal point between the condenser and objective, and contains local phase gradients $m$ and $n$ in the horizontal and vertical directions, the effect of which is to tilt the beam slightly after it passes through the sample, by an amount that is proportional to $m$ and $n$. When using a smooth aperture, the intensity difference between the tilted and un-tilted beam is negligible and no change is visible in the image. However, if an aperture has a sharp edge then a small tilt can have a significant effect on the intensity of that point in the image. See for example $P_c$ in figure~\ref{fig:Setup} where the left half the beam has been blocked before the condenser producing a sharp vertical edge. A positive gradient would shift the beam in one direction, and a negative gradient would shift the beam in the other direction resulting in, for example, the left side of a phase object appearing light, and the right side dark. If the aperture were flipped such that the right half of the beam is blocked, then the highlighting would be inverted and the right side of the object would appear light, and the left side dark. This effect is apparent in figures~\ref{fig2}~and~\ref{fig:Contrast}. } 

Asymmetric images formed from complimentary apertures can in turn be used to create DPC images $I_{DPC}$, using the following~\cite{Mehta2009}:
\begin{equation}
    I_{DPC}=2\frac{I_R-I_L}{I_R+I_L}
    \label{eq:DPC}
\end{equation}
Where $I_L$ and $I_R$ are images formed by asymmetric illumination from the left and right side respectively. Here a right-left asymmetric example is used for clarity, but any appropriate aperture could be used.

In this paper, we show that, by making appropriate post-selective measurements on correlated photon pairs, we can apply a digital aperture $P_d$ which is exactly equivalent to a physical aperture $P_c$. This equivalence is shown in figure~\ref{fig:Setup}. \COR{The contrast and resolution that can be achieved with with this digital aperture is no better than could be achieved with a physical aperture, but rather the advantage in this technique comes from the fact that $P_d$ can be dynamically reconfigured in post-processing after the image is taken. To quantify the phase gradient, the numerical apertures and phase gradients must all be appropriately calibrated and normalised. In this proof-of principle demonstration, the stability and repeatability of the microscope was insufficient to perform this calibration. Instead we show qualitatively that phase gradients are highlighted in the output image.} 

Our experimental setup centres on three elements: a time-tagging single photon sensitive camera, a phase target, and a correlated photon pair source. These components are discussed sequentially below. The camera used is the TPX3CAM~\cite{Nomerotski2019}, which is an event-driven camera with $256\times256$ pixels, each $55\times55$\,$\mu m$ in size. The camera is not single photon sensitive so a fast image intensifier is appended. Because the intensifier produces many photons for each incident photon, cluster identification and centroiding is required to identify a single event. After these corrections (see~\cite{Ianzano2020} for details), a 3D data set emerges containing the ($x,y$) coordinate and arrival time of every photon arriving at the intensifier. The total detection efficiency is 7\% and is approximately uniform across the camera. The arrival-time jitter is dominated by effects related to the intensifier, and in this case is 16\,ns FWHM~\cite{Zhang2020,vidyapin_characterisation_2022}. \COR{The imaging optics produce a $40\times$ magnification of the image, with a $\sim$40\,$\mu$m field of view (FoV). Due to the need of using a single camera to image both photons, the signal beam was demagnified such that it only illuminates a $\sim80\times80$\,pixel area of camera. In the ideal situation where two cameras are used, one for each photon, the imaging optics could be changed such that the FoV could be extended by a factor of $\sim3$ for the same resolution. Or, equivalently, the resolution could be increased by a factor of $\sim3$ for the same FOV.}

Two types of sample were used, both exhibiting features on the micron scale. First two standardized phase targets, with etching depths of 150\,nm and 350\,nm (Fig.~\ref{fig2}(a)\,). Second, an unstained cell removed from the inside of a human cheek using a swab and transferred to a microscope slide (Fig.~\ref{fig2}(b)\,). The data shown in all figures was obtained in a 250\,s exposure time. Exposure times as short as 1\,s can produce visible images, but longer data sets produce better quality images and allow more restrictive digital apertures in post-processing. 

The photon pairs are generated by SPDC in a 1\,mm thick type II periodically poled potassium titanyl phosphate (ppKTP) crystal, pumped by a $\sim$100\,mW 405\,nm continuous wave (CW) diode laser. The emitted photons have orthogonal polarization and are correlated in position and anti-correlated in their momentum. The photons are first separated with a polarizing beam splitter (PBS) and in the path of the horizontally polarized `signal' photons, the plane of the crystal is first imaged onto the target, and then re-imaged onto the camera.  Meanwhile, in the path of the vertically polarized `idler' photons, the fourier plane of the crystal is imaged onto a different part of the same camera. See Fig.~\ref{fig:Setup} for details. 

The first step in the analysis is to identify coincident pairs of photons in the image. A coincidence window of 20\,ns is chosen due to the temporal jitter of the apparatus. If two photons are incident on the camera, one in the near field and one in the far field, and their arrival time difference is $<20$\,ns, then they are considered as a pair. All non-paired events are discarded. The result is a `coincidence image' made up entirely of pairs of photons. Knowing that the photon pairs are momentum anti-correlated, the illumination angle of each photon on the target can be deduced based on where its partner photon was detected in the far-field. The analysis proceeds by making a post-selection in the far-field such that only a subset of the photon pairs are used to form an image. In this way, we can control the illumination angles which are used to form the image, this is the direct analog of placing a physical aperture before the condenser lens so we consider this to be a digital aperture.

Figure~\ref{fig2} shows how asymmetric and DPC images can be formed from this data, and illustrates the advantages compared to full-field illumination. Figures~\ref{fig2}(a1) and (b1) show typical raw images with the far-field (angle) and near-field (position) modes of the 1951 USAF resolution phase target and an unstained cheek cell, respectively. In the near-field we see a low-contrast image of the sample. In Fig.~\ref{fig2}(a2) and (b2), 4 different hemispherical digital apertures have been applied to the far-field and coincidence images corresponding to each aperture are shown. \COR{It is apparent that the digital apertures have altered the appearance of the images, particularly the edge gradients are highlighted}. Furthermore, there are clear differences between the images formed using apertures at different angles, revealing different sections of the target. \COR{This illustrates an interesting feature of this scheme: the aperture can be manipulated in post-processing after the image is taken, so one or more different apertures can be used to highlight particular features.} Figure~\ref{fig2}(a3) and (b3) shows how the asymmetric images can be used to form a DPC image according to Eq.~(\ref{eq:DPC}). From the DPC image of the resolution target, we see that the phase gradients parallel to the edge of the digital aperture produces the highest contrast.

In Fig.~\ref{fig:Contrast}, a raw image  is compared to asymmetric images formed from the right (RHS) and left (LHS) for the phase resolution targets with an etching depth of Fig.~\ref{fig:Contrast}(a) 350\,nm and Fig.~\ref{fig:Contrast}(b) 150\,nm . To quantify these differences, we integrate the photon counts across the same 3 lines in the image for each selection. As can be seen, the left-hand side and right-hand side images shows higher contrast level compared to the raw image. The left and right-hand side images having shifted contrast patterns as opposite sides of the lines are highlighted. As expected, this comparison not only indicates that the asymmetric selections improves the contrast levels, but that the highlighted areas shift as the angle of illumination changes. We use the following expression to quantify the visibility of the lines: 
\begin{equation}
    V=\frac{\bar{I}_{\text{max}}-\bar{I}_{\text{min}}}{\bar{I}_{\text{max}}+\bar{I}_{\text{min}}},
\end{equation}
where $\bar{I}_{\text{max}}$ is the average maximum intensity of the 3 lines and $\bar{I}_{\text{min}}$ is the average mimimum intensity in between the 3 lines. Using this we get for the 350\,nm target $V_{\text{LHS}} = 16$\%, $V_{\text{RHS}} = 16$\%, $V_{\text{raw}} = 7.4$\% and for the 150\,nm target $V_{\text{LHS}} = 7.8$\%, $V_{\text{RHS}} = 9.4$\%, $V_{\text{raw}} = 1.9$\%.

In conclusion, \COR{we have proposed and demonstrated a technique using photon pairs produced by SPDC which is sensitive to local phase gradients in a sample}. Digital apertures in the Fourier plane of any shape can be applied in post-processing taking advantage of the spatial-temporal quantum correlations. The technique works for unstained or phase objects, and offers freedom to the experimentalist in choosing the shape of the Fourier-filter in post-processing. \COR{Noise, in the form of electrical dark current or background light, is uncorrelated. Conversely, the arrival time of photon pairs produced by SPDC is highly correlated. Accordingly, by selecting only pairs of events that arrive at the same time and discarding all single events, these correlations can be used to suppress undesirable noise in the image~\cite{England2019}. These noise-suppression features, enabled by the quantum correlations in the SPDC source, mean that our technique can operate with just picowatts of illuminating power and will be well-suited to non-invasive imaging applications. }


In the future, such a post-processing based image system could allow bright-field, dark-field, phase contrast, and 3D images~\cite{Zhang2022} to all be drawn from a single data set. \COR{This approach is complimentary to existing depth-sensitive microscopy techniques based on Hong-Ou-Mandel (HOM) interference of photon pairs~\cite{ndagano_quantum_2022,torre_sub-mum_2022}. These use two-photon interference to measure absolute path length differences in reflective samples, whereas our technique employs position-momentum correlations in transmissive samples to highlight local phase gradients.  } 

The main limiting factor in our demonstration is the detection technology, in particular the detection efficiency of our camera which only 7\%~\cite{vidyapin_characterisation_2022}. Since the required data acquisition time scales quadratically for coincidence measurements, improvements in efficiency would greatly increase the speed of this technique. For example, a camera with $50\%$ efficiency would acquire similar images in just 5 seconds. Furthermore, the timing resolution of this camera is $\sim 10$\,ns and is a limiting factor to the noise suppression gained through the temporal correlations. With the timing resolution improved to 0.1\,ns, the signal to noise ratio could be increased by a factor of 10. Single photon event camera technologies satisfying these specifications are already under development~\cite{Morimoto2020,Canon}, so integration times of less than 1\,s could be possible in the near future. Higher detector efficiency and resolution will also unlock new functionalities. For example, digital apertures that result in high loss (such as a high-pass filter for dark-field imaging) cannot be effectively implemented with this setup, but could be possible with improved detectors.

\begin{acknowledgments}
The authors are grateful to Andrei Nomerotski, Antony Orth, Adrian Pegoraro, Andrew Ridsdale, Victor Vidyapin, Denis Guay, and Doug Moffatt for stimulating discussions and technical support. This work was partly supported by Defence Research and Development Canada and the National Research Council's Quantum Sensors Challenge Program. 
\end{acknowledgments}

\section*{Data Availability}
The data that support the findings of this study are available from the corresponding author upon reasonable request.

\section*{Author Declarations}
The authors have no conflicts to disclose.

\
\section*{References}
\nocite{*}
\bibliography{Asymmetric}

\begin{thebibliography}{39}%
\makeatletter
\providecommand \@ifxundefined [1]{%
 \@ifx{#1\undefined}
}%
\providecommand \@ifnum [1]{%
 \ifnum #1\expandafter \@firstoftwo
 \else \expandafter \@secondoftwo
 \fi
}%
\providecommand \@ifx [1]{%
 \ifx #1\expandafter \@firstoftwo
 \else \expandafter \@secondoftwo
 \fi
}%
\providecommand \natexlab [1]{#1}%
\providecommand \enquote  [1]{``#1''}%
\providecommand \bibnamefont  [1]{#1}%
\providecommand \bibfnamefont [1]{#1}%
\providecommand \citenamefont [1]{#1}%
\providecommand \href@noop [0]{\@secondoftwo}%
\providecommand \href [0]{\begingroup \@sanitize@url \@href}%
\providecommand \@href[1]{\@@startlink{#1}\@@href}%
\providecommand \@@href[1]{\endgroup#1\@@endlink}%
\providecommand \@sanitize@url [0]{\catcode `\\12\catcode `\$12\catcode
  `\&12\catcode `\#12\catcode `\^12\catcode `\_12\catcode `\%12\relax}%
\providecommand \@@startlink[1]{}%
\providecommand \@@endlink[0]{}%
\providecommand \url  [0]{\begingroup\@sanitize@url \@url }%
\providecommand \@url [1]{\endgroup\@href {#1}{\urlprefix }}%
\providecommand \urlprefix  [0]{URL }%
\providecommand \Eprint [0]{\href }%
\providecommand \doibase [0]{http://dx.doi.org/}%
\providecommand \selectlanguage [0]{\@gobble}%
\providecommand \bibinfo  [0]{\@secondoftwo}%
\providecommand \bibfield  [0]{\@secondoftwo}%
\providecommand \translation [1]{[#1]}%
\providecommand \BibitemOpen [0]{}%
\providecommand \bibitemStop [0]{}%
\providecommand \bibitemNoStop [0]{.\EOS\space}%
\providecommand \EOS [0]{\spacefactor3000\relax}%
\providecommand \BibitemShut  [1]{\csname bibitem#1\endcsname}%
\let\auto@bib@innerbib\@empty
\bibitem [{\citenamefont {Smith}(2014)}]{Smith2014}%
  \BibitemOpen
  \bibfield  {author} {\bibinfo {author} {\bibfnamefont {A.~M.}\ \bibnamefont
  {Smith}},\ }\bibfield  {title} {\enquote {\bibinfo {title} {From sight to
  light},}\ }in\ \href@noop {} {\emph {\bibinfo {booktitle} {From Sight to
  Light}}}\ (\bibinfo  {publisher} {University of Chicago Press},\ \bibinfo
  {year} {2014})\BibitemShut {NoStop}%
\bibitem [{\citenamefont {Defienne}\ \emph {et~al.}(2021)\citenamefont
  {Defienne}, \citenamefont {Ndagano}, \citenamefont {Lyons},\ and\
  \citenamefont {Faccio}}]{Defienne2021}%
  \BibitemOpen
  \bibfield  {author} {\bibinfo {author} {\bibfnamefont {H.}~\bibnamefont
  {Defienne}}, \bibinfo {author} {\bibfnamefont {B.}~\bibnamefont {Ndagano}},
  \bibinfo {author} {\bibfnamefont {A.}~\bibnamefont {Lyons}}, \ and\ \bibinfo
  {author} {\bibfnamefont {D.}~\bibnamefont {Faccio}},\ }\bibfield  {title}
  {\enquote {\bibinfo {title} {Polarization entanglement-enabled quantum
  holography},}\ }\href@noop {} {\bibfield  {journal} {\bibinfo  {journal}
  {Nature Physics}\ }\textbf {\bibinfo {volume} {17}},\ \bibinfo {pages}
  {591--597} (\bibinfo {year} {2021})}\BibitemShut {NoStop}%
\bibitem [{\citenamefont {Casacio}\ \emph {et~al.}(2021)\citenamefont
  {Casacio}, \citenamefont {Madsen}, \citenamefont {Terrasson}, \citenamefont
  {Waleed}, \citenamefont {Barnscheidt}, \citenamefont {Hage}, \citenamefont
  {Taylor},\ and\ \citenamefont {Bowen}}]{Casacio2021}%
  \BibitemOpen
  \bibfield  {author} {\bibinfo {author} {\bibfnamefont {C.~A.}\ \bibnamefont
  {Casacio}}, \bibinfo {author} {\bibfnamefont {L.~S.}\ \bibnamefont {Madsen}},
  \bibinfo {author} {\bibfnamefont {A.}~\bibnamefont {Terrasson}}, \bibinfo
  {author} {\bibfnamefont {M.}~\bibnamefont {Waleed}}, \bibinfo {author}
  {\bibfnamefont {K.}~\bibnamefont {Barnscheidt}}, \bibinfo {author}
  {\bibfnamefont {B.}~\bibnamefont {Hage}}, \bibinfo {author} {\bibfnamefont
  {M.~A.}\ \bibnamefont {Taylor}}, \ and\ \bibinfo {author} {\bibfnamefont
  {W.~P.}\ \bibnamefont {Bowen}},\ }\bibfield  {title} {\enquote {\bibinfo
  {title} {Quantum-enhanced nonlinear microscopy},}\ }\href@noop {} {\bibfield
  {journal} {\bibinfo  {journal} {Nature}\ }\textbf {\bibinfo {volume} {594}},\
  \bibinfo {pages} {201--206} (\bibinfo {year} {2021})}\BibitemShut {NoStop}%
\bibitem [{\citenamefont {D'Angelo}, \citenamefont {Chekhova},\ and\
  \citenamefont {Shih}(2001)}]{dangelo_two-photon_2001}%
  \BibitemOpen
  \bibfield  {author} {\bibinfo {author} {\bibfnamefont {M.}~\bibnamefont
  {D'Angelo}}, \bibinfo {author} {\bibfnamefont {M.~V.}\ \bibnamefont
  {Chekhova}}, \ and\ \bibinfo {author} {\bibfnamefont {Y.}~\bibnamefont
  {Shih}},\ }\bibfield  {title} {\enquote {\bibinfo {title} {Two-{Photon}
  {Diffraction} and {Quantum} {Lithography}},}\ }\href {\doibase
  10.1103/PhysRevLett.87.013602} {\bibfield  {journal} {\bibinfo  {journal}
  {Physical Review Letters}\ }\textbf {\bibinfo {volume} {87}},\ \bibinfo
  {pages} {013602} (\bibinfo {year} {2001})},\ \bibinfo {note} {publisher:
  American Physical Society}\BibitemShut {NoStop}%
\bibitem [{\citenamefont {Israel}, \citenamefont {Rosen},\ and\ \citenamefont
  {Silberberg}(2014)}]{israel_supersensitive_2014}%
  \BibitemOpen
  \bibfield  {author} {\bibinfo {author} {\bibfnamefont {Y.}~\bibnamefont
  {Israel}}, \bibinfo {author} {\bibfnamefont {S.}~\bibnamefont {Rosen}}, \
  and\ \bibinfo {author} {\bibfnamefont {Y.}~\bibnamefont {Silberberg}},\
  }\bibfield  {title} {\enquote {\bibinfo {title} {Supersensitive
  {Polarization} {Microscopy} {Using} {NOON} {States} of {Light}},}\ }\href
  {\doibase 10.1103/PhysRevLett.112.103604} {\bibfield  {journal} {\bibinfo
  {journal} {Physical Review Letters}\ }\textbf {\bibinfo {volume} {112}},\
  \bibinfo {pages} {103604} (\bibinfo {year} {2014})},\ \bibinfo {note}
  {publisher: American Physical Society}\BibitemShut {NoStop}%
\bibitem [{\citenamefont {D’Angelo}\ \emph {et~al.}(2005)\citenamefont
  {D’Angelo}, \citenamefont {Valencia}, \citenamefont {Rubin},\ and\
  \citenamefont {Shih}}]{dangelo_resolution_2005}%
  \BibitemOpen
  \bibfield  {author} {\bibinfo {author} {\bibfnamefont {M.}~\bibnamefont
  {D’Angelo}}, \bibinfo {author} {\bibfnamefont {A.}~\bibnamefont
  {Valencia}}, \bibinfo {author} {\bibfnamefont {M.~H.}\ \bibnamefont {Rubin}},
  \ and\ \bibinfo {author} {\bibfnamefont {Y.}~\bibnamefont {Shih}},\
  }\bibfield  {title} {\enquote {\bibinfo {title} {Resolution of quantum and
  classical ghost imaging},}\ }\href {\doibase 10.1103/PhysRevA.72.013810}
  {\bibfield  {journal} {\bibinfo  {journal} {Physical Review A}\ }\textbf
  {\bibinfo {volume} {72}},\ \bibinfo {pages} {013810} (\bibinfo {year}
  {2005})},\ \bibinfo {note} {publisher: American Physical Society}\BibitemShut
  {NoStop}%
\bibitem [{\citenamefont {Padgett}\ and\ \citenamefont
  {Boyd}(2017)}]{Padgett2017}%
  \BibitemOpen
  \bibfield  {author} {\bibinfo {author} {\bibfnamefont {M.~J.}\ \bibnamefont
  {Padgett}}\ and\ \bibinfo {author} {\bibfnamefont {R.~W.}\ \bibnamefont
  {Boyd}},\ }\bibfield  {title} {\enquote {\bibinfo {title} {An introduction to
  ghost imaging: quantum and classical},}\ }\href@noop {} {\bibfield  {journal}
  {\bibinfo  {journal} {Philosophical Transactions of the Royal Society A:
  Mathematical, Physical and Engineering Sciences}\ }\textbf {\bibinfo {volume}
  {375}},\ \bibinfo {pages} {20160233} (\bibinfo {year} {2017})}\BibitemShut
  {NoStop}%
\bibitem [{\citenamefont {Gregory}\ \emph {et~al.}(2020)\citenamefont
  {Gregory}, \citenamefont {Moreau}, \citenamefont {Toninelli},\ and\
  \citenamefont {Padgett}}]{gregory_imaging_2020}%
  \BibitemOpen
  \bibfield  {author} {\bibinfo {author} {\bibfnamefont {T.}~\bibnamefont
  {Gregory}}, \bibinfo {author} {\bibfnamefont {P.-A.}\ \bibnamefont {Moreau}},
  \bibinfo {author} {\bibfnamefont {E.}~\bibnamefont {Toninelli}}, \ and\
  \bibinfo {author} {\bibfnamefont {M.~J.}\ \bibnamefont {Padgett}},\
  }\bibfield  {title} {\enquote {\bibinfo {title} {Imaging through noise with
  quantum illumination},}\ }\href {\doibase 10.1126/sciadv.aay2652} {\bibfield
  {journal} {\bibinfo  {journal} {Science Advances}\ }\textbf {\bibinfo
  {volume} {6}},\ \bibinfo {pages} {eaay2652} (\bibinfo {year}
  {2020})}\BibitemShut {NoStop}%
\bibitem [{\citenamefont {Brida}, \citenamefont {Genovese},\ and\ \citenamefont
  {Berchera}(2010)}]{Brida2010}%
  \BibitemOpen
  \bibfield  {author} {\bibinfo {author} {\bibfnamefont {G.}~\bibnamefont
  {Brida}}, \bibinfo {author} {\bibfnamefont {M.}~\bibnamefont {Genovese}}, \
  and\ \bibinfo {author} {\bibfnamefont {I.~R.}\ \bibnamefont {Berchera}},\
  }\bibfield  {title} {\enquote {\bibinfo {title} {Experimental realization of
  sub-shot-noise quantum imaging},}\ }\href@noop {} {\bibfield  {journal}
  {\bibinfo  {journal} {Nature Photonics}\ }\textbf {\bibinfo {volume} {4}},\
  \bibinfo {pages} {227--230} (\bibinfo {year} {2010})}\BibitemShut {NoStop}%
\bibitem [{\citenamefont {Lemos}\ \emph {et~al.}(2014)\citenamefont {Lemos},
  \citenamefont {Borish}, \citenamefont {Cole}, \citenamefont {Ramelow},
  \citenamefont {Lapkiewicz},\ and\ \citenamefont {Zeilinger}}]{Lemos2014}%
  \BibitemOpen
  \bibfield  {author} {\bibinfo {author} {\bibfnamefont {G.~B.}\ \bibnamefont
  {Lemos}}, \bibinfo {author} {\bibfnamefont {V.}~\bibnamefont {Borish}},
  \bibinfo {author} {\bibfnamefont {G.~D.}\ \bibnamefont {Cole}}, \bibinfo
  {author} {\bibfnamefont {S.}~\bibnamefont {Ramelow}}, \bibinfo {author}
  {\bibfnamefont {R.}~\bibnamefont {Lapkiewicz}}, \ and\ \bibinfo {author}
  {\bibfnamefont {A.}~\bibnamefont {Zeilinger}},\ }\bibfield  {title} {\enquote
  {\bibinfo {title} {Quantum imaging with undetected photons},}\ }\href@noop {}
  {\bibfield  {journal} {\bibinfo  {journal} {Nature}\ }\textbf {\bibinfo
  {volume} {512}},\ \bibinfo {pages} {409--412} (\bibinfo {year}
  {2014})}\BibitemShut {NoStop}%
\bibitem [{\citenamefont {Zhang}, \citenamefont {England},\ and\ \citenamefont
  {Sussman}(2022)}]{zhang_snapshot_2022}%
  \BibitemOpen
  \bibfield  {author} {\bibinfo {author} {\bibfnamefont {Y.}~\bibnamefont
  {Zhang}}, \bibinfo {author} {\bibfnamefont {D.}~\bibnamefont {England}}, \
  and\ \bibinfo {author} {\bibfnamefont {B.}~\bibnamefont {Sussman}},\ }\href
  {\doibase 10.48550/arXiv.2204.05984} {\enquote {\bibinfo {title} {Snapshot
  hyperspectral imaging with quantum correlated photons},}\ } (\bibinfo {year}
  {2022}),\ \bibinfo {note} {arXiv:2204.05984 [physics,
  physics:quant-ph]}\BibitemShut {NoStop}%
\bibitem [{\citenamefont {Genovese}(2016)}]{Genovese2016}%
  \BibitemOpen
  \bibfield  {author} {\bibinfo {author} {\bibfnamefont {M.}~\bibnamefont
  {Genovese}},\ }\bibfield  {title} {\enquote {\bibinfo {title} {Real
  applications of quantum imaging},}\ }\href@noop {} {\bibfield  {journal}
  {\bibinfo  {journal} {Journal of Optics}\ }\textbf {\bibinfo {volume} {18}},\
  \bibinfo {pages} {073002} (\bibinfo {year} {2016})}\BibitemShut {NoStop}%
\bibitem [{\citenamefont {Zhang}\ \emph {et~al.}(2022)\citenamefont {Zhang},
  \citenamefont {Orth}, \citenamefont {England},\ and\ \citenamefont
  {Sussman}}]{Zhang2022}%
  \BibitemOpen
  \bibfield  {author} {\bibinfo {author} {\bibfnamefont {Y.}~\bibnamefont
  {Zhang}}, \bibinfo {author} {\bibfnamefont {A.}~\bibnamefont {Orth}},
  \bibinfo {author} {\bibfnamefont {D.}~\bibnamefont {England}}, \ and\
  \bibinfo {author} {\bibfnamefont {B.}~\bibnamefont {Sussman}},\ }\bibfield
  {title} {\enquote {\bibinfo {title} {Ray tracing with quantum correlated
  photons to image a three-dimensional scene},}\ }\href {\doibase
  10.1103/PhysRevA.105.L011701} {\bibfield  {journal} {\bibinfo  {journal}
  {Phys. Rev. A}\ }\textbf {\bibinfo {volume} {105}},\ \bibinfo {pages}
  {L011701} (\bibinfo {year} {2022})}\BibitemShut {NoStop}%
\bibitem [{\citenamefont {Maurer}\ \emph {et~al.}(2011)\citenamefont {Maurer},
  \citenamefont {Jesacher}, \citenamefont {Bernet},\ and\ \citenamefont
  {Ritsch-Marte}}]{Maurer2011}%
  \BibitemOpen
  \bibfield  {author} {\bibinfo {author} {\bibfnamefont {C.}~\bibnamefont
  {Maurer}}, \bibinfo {author} {\bibfnamefont {A.}~\bibnamefont {Jesacher}},
  \bibinfo {author} {\bibfnamefont {S.}~\bibnamefont {Bernet}}, \ and\ \bibinfo
  {author} {\bibfnamefont {M.}~\bibnamefont {Ritsch-Marte}},\ }\bibfield
  {title} {\enquote {\bibinfo {title} {What spatial light modulators can do for
  optical microscopy},}\ }\href@noop {} {\bibfield  {journal} {\bibinfo
  {journal} {Laser \& Photonics Reviews}\ }\textbf {\bibinfo {volume} {5}},\
  \bibinfo {pages} {81--101} (\bibinfo {year} {2011})}\BibitemShut {NoStop}%
\bibitem [{\citenamefont {Liu}\ \emph {et~al.}(2014)\citenamefont {Liu},
  \citenamefont {Tian}, \citenamefont {Liu},\ and\ \citenamefont
  {Waller}}]{Liu2014}%
  \BibitemOpen
  \bibfield  {author} {\bibinfo {author} {\bibfnamefont {Z.}~\bibnamefont
  {Liu}}, \bibinfo {author} {\bibfnamefont {L.}~\bibnamefont {Tian}}, \bibinfo
  {author} {\bibfnamefont {S.}~\bibnamefont {Liu}}, \ and\ \bibinfo {author}
  {\bibfnamefont {L.}~\bibnamefont {Waller}},\ }\bibfield  {title} {\enquote
  {\bibinfo {title} {{Real-time brightfield, darkfield, and phase contrast
  imaging in a light-emitting diode array microscope}},}\ }\href {\doibase
  10.1117/1.JBO.19.10.106002} {\bibfield  {journal} {\bibinfo  {journal}
  {Journal of Biomedical Optics}\ }\textbf {\bibinfo {volume} {19}},\ \bibinfo
  {pages} {106002} (\bibinfo {year} {2014})}\BibitemShut {NoStop}%
\bibitem [{\citenamefont {Tian}, \citenamefont {Wang},\ and\ \citenamefont
  {Waller}(2014)}]{Tian2014}%
  \BibitemOpen
  \bibfield  {author} {\bibinfo {author} {\bibfnamefont {L.}~\bibnamefont
  {Tian}}, \bibinfo {author} {\bibfnamefont {J.}~\bibnamefont {Wang}}, \ and\
  \bibinfo {author} {\bibfnamefont {L.}~\bibnamefont {Waller}},\ }\bibfield
  {title} {\enquote {\bibinfo {title} {3d differential phase-contrast
  microscopy with computational illumination using an led array},}\ }\href@noop
  {} {\bibfield  {journal} {\bibinfo  {journal} {Optics letters}\ }\textbf
  {\bibinfo {volume} {39}},\ \bibinfo {pages} {1326--1329} (\bibinfo {year}
  {2014})}\BibitemShut {NoStop}%
\bibitem [{\citenamefont {Tian}\ and\ \citenamefont {Waller}(2015)}]{Tian2015}%
  \BibitemOpen
  \bibfield  {author} {\bibinfo {author} {\bibfnamefont {L.}~\bibnamefont
  {Tian}}\ and\ \bibinfo {author} {\bibfnamefont {L.}~\bibnamefont {Waller}},\
  }\bibfield  {title} {\enquote {\bibinfo {title} {Quantitative differential
  phase contrast imaging in an led array microscope},}\ }\href@noop {}
  {\bibfield  {journal} {\bibinfo  {journal} {Optics express}\ }\textbf
  {\bibinfo {volume} {23}},\ \bibinfo {pages} {11394--11403} (\bibinfo {year}
  {2015})}\BibitemShut {NoStop}%
\bibitem [{\citenamefont {Lee}\ \emph {et~al.}(2015)\citenamefont {Lee},
  \citenamefont {Ryu}, \citenamefont {Kim}, \citenamefont {Jung},\ and\
  \citenamefont {Joo}}]{Lee2015}%
  \BibitemOpen
  \bibfield  {author} {\bibinfo {author} {\bibfnamefont {D.}~\bibnamefont
  {Lee}}, \bibinfo {author} {\bibfnamefont {S.}~\bibnamefont {Ryu}}, \bibinfo
  {author} {\bibfnamefont {U.}~\bibnamefont {Kim}}, \bibinfo {author}
  {\bibfnamefont {D.}~\bibnamefont {Jung}}, \ and\ \bibinfo {author}
  {\bibfnamefont {C.}~\bibnamefont {Joo}},\ }\bibfield  {title} {\enquote
  {\bibinfo {title} {Color-coded led microscopy for multi-contrast and
  quantitative phase-gradient imaging},}\ }\href {\doibase
  10.1364/BOE.6.004912} {\bibfield  {journal} {\bibinfo  {journal} {Biomed.
  Opt. Express}\ }\textbf {\bibinfo {volume} {6}},\ \bibinfo {pages}
  {4912--4922} (\bibinfo {year} {2015})}\BibitemShut {NoStop}%
\bibitem [{\citenamefont {Jung}\ \emph {et~al.}(2017)\citenamefont {Jung},
  \citenamefont {Choi}, \citenamefont {Kim}, \citenamefont {Ryu}, \citenamefont
  {Lee}, \citenamefont {Lee},\ and\ \citenamefont {Joo}}]{Jung2017}%
  \BibitemOpen
  \bibfield  {author} {\bibinfo {author} {\bibfnamefont {D.}~\bibnamefont
  {Jung}}, \bibinfo {author} {\bibfnamefont {J.-H.}\ \bibnamefont {Choi}},
  \bibinfo {author} {\bibfnamefont {S.}~\bibnamefont {Kim}}, \bibinfo {author}
  {\bibfnamefont {S.}~\bibnamefont {Ryu}}, \bibinfo {author} {\bibfnamefont
  {W.}~\bibnamefont {Lee}}, \bibinfo {author} {\bibfnamefont {J.-S.}\
  \bibnamefont {Lee}}, \ and\ \bibinfo {author} {\bibfnamefont
  {C.}~\bibnamefont {Joo}},\ }\bibfield  {title} {\enquote {\bibinfo {title}
  {Smartphone-based multi-contrast microscope using color-multiplexed
  illumination},}\ }\href {\doibase 10.1038/s41598-017-07703-w} {\bibfield
  {journal} {\bibinfo  {journal} {Scientific Reports}\ }\textbf {\bibinfo
  {volume} {7}},\ \bibinfo {pages} {7564} (\bibinfo {year} {2017})}\BibitemShut
  {NoStop}%
\bibitem [{\citenamefont {Goedhart}\ \emph {et~al.}(2007)\citenamefont
  {Goedhart}, \citenamefont {Khalilzada}, \citenamefont {Bezemer},
  \citenamefont {Merza},\ and\ \citenamefont {Ince}}]{Goedhart2007}%
  \BibitemOpen
  \bibfield  {author} {\bibinfo {author} {\bibfnamefont {P.}~\bibnamefont
  {Goedhart}}, \bibinfo {author} {\bibfnamefont {M.}~\bibnamefont
  {Khalilzada}}, \bibinfo {author} {\bibfnamefont {R.}~\bibnamefont {Bezemer}},
  \bibinfo {author} {\bibfnamefont {J.}~\bibnamefont {Merza}}, \ and\ \bibinfo
  {author} {\bibfnamefont {C.}~\bibnamefont {Ince}},\ }\bibfield  {title}
  {\enquote {\bibinfo {title} {Sidestream dark field (sdf) imaging: a novel
  stroboscopic led ring-based imaging modality for clinical assessment of the
  microcirculation.}}\ }\href@noop {} {\bibfield  {journal} {\bibinfo
  {journal} {Optics express}\ }\textbf {\bibinfo {volume} {15}},\ \bibinfo
  {pages} {15101--15114} (\bibinfo {year} {2007})}\BibitemShut {NoStop}%
\bibitem [{\citenamefont {Zheng}, \citenamefont {Kolner},\ and\ \citenamefont
  {Yang}(2011)}]{Zheng2011}%
  \BibitemOpen
  \bibfield  {author} {\bibinfo {author} {\bibfnamefont {G.}~\bibnamefont
  {Zheng}}, \bibinfo {author} {\bibfnamefont {C.}~\bibnamefont {Kolner}}, \
  and\ \bibinfo {author} {\bibfnamefont {C.}~\bibnamefont {Yang}},\ }\bibfield
  {title} {\enquote {\bibinfo {title} {Microscopy refocusing and dark-field
  imaging by using a simple led array},}\ }\href@noop {} {\bibfield  {journal}
  {\bibinfo  {journal} {Optics letters}\ }\textbf {\bibinfo {volume} {36}},\
  \bibinfo {pages} {3987--3989} (\bibinfo {year} {2011})}\BibitemShut {NoStop}%
\bibitem [{\citenamefont {Horstmeyer}\ \emph {et~al.}(2016)\citenamefont
  {Horstmeyer}, \citenamefont {Chung}, \citenamefont {Ou}, \citenamefont
  {Zheng},\ and\ \citenamefont {Yang}}]{Horstmeyer2016}%
  \BibitemOpen
  \bibfield  {author} {\bibinfo {author} {\bibfnamefont {R.}~\bibnamefont
  {Horstmeyer}}, \bibinfo {author} {\bibfnamefont {J.}~\bibnamefont {Chung}},
  \bibinfo {author} {\bibfnamefont {X.}~\bibnamefont {Ou}}, \bibinfo {author}
  {\bibfnamefont {G.}~\bibnamefont {Zheng}}, \ and\ \bibinfo {author}
  {\bibfnamefont {C.}~\bibnamefont {Yang}},\ }\bibfield  {title} {\enquote
  {\bibinfo {title} {Diffraction tomography with fourier ptychography},}\
  }\href@noop {} {\bibfield  {journal} {\bibinfo  {journal} {Optica}\ }\textbf
  {\bibinfo {volume} {3}},\ \bibinfo {pages} {827--835} (\bibinfo {year}
  {2016})}\BibitemShut {NoStop}%
\bibitem [{\citenamefont {Ma}\ \emph {et~al.}(2015)\citenamefont {Ma},
  \citenamefont {Liu}, \citenamefont {Tian}, \citenamefont {Dai},\ and\
  \citenamefont {Waller}}]{Ma2015}%
  \BibitemOpen
  \bibfield  {author} {\bibinfo {author} {\bibfnamefont {C.}~\bibnamefont
  {Ma}}, \bibinfo {author} {\bibfnamefont {Z.}~\bibnamefont {Liu}}, \bibinfo
  {author} {\bibfnamefont {L.}~\bibnamefont {Tian}}, \bibinfo {author}
  {\bibfnamefont {Q.}~\bibnamefont {Dai}}, \ and\ \bibinfo {author}
  {\bibfnamefont {L.}~\bibnamefont {Waller}},\ }\bibfield  {title} {\enquote
  {\bibinfo {title} {Motion deblurring with temporally coded illumination in an
  led array microscope},}\ }\href@noop {} {\bibfield  {journal} {\bibinfo
  {journal} {Optics letters}\ }\textbf {\bibinfo {volume} {40}},\ \bibinfo
  {pages} {2281--2284} (\bibinfo {year} {2015})}\BibitemShut {NoStop}%
\bibitem [{\citenamefont {Mehta}\ and\ \citenamefont
  {Sheppard}(2009)}]{Mehta2009}%
  \BibitemOpen
  \bibfield  {author} {\bibinfo {author} {\bibfnamefont {S.~B.}\ \bibnamefont
  {Mehta}}\ and\ \bibinfo {author} {\bibfnamefont {C.~J.}\ \bibnamefont
  {Sheppard}},\ }\bibfield  {title} {\enquote {\bibinfo {title} {Quantitative
  phase-gradient imaging at high resolution with asymmetric illumination-based
  differential phase contrast},}\ }\href@noop {} {\bibfield  {journal}
  {\bibinfo  {journal} {Optics letters}\ }\textbf {\bibinfo {volume} {34}},\
  \bibinfo {pages} {1924--1926} (\bibinfo {year} {2009})}\BibitemShut {NoStop}%
\bibitem [{\citenamefont {Hamilton}\ and\ \citenamefont
  {Sheppard}(1984)}]{Hamilton1984}%
  \BibitemOpen
  \bibfield  {author} {\bibinfo {author} {\bibfnamefont {D.}~\bibnamefont
  {Hamilton}}\ and\ \bibinfo {author} {\bibfnamefont {C.}~\bibnamefont
  {Sheppard}},\ }\bibfield  {title} {\enquote {\bibinfo {title} {Differential
  phase contrast in scanning optical microscopy},}\ }\href@noop {} {\bibfield
  {journal} {\bibinfo  {journal} {Journal of microscopy}\ }\textbf {\bibinfo
  {volume} {133}},\ \bibinfo {pages} {27--39} (\bibinfo {year}
  {1984})}\BibitemShut {NoStop}%
\bibitem [{\citenamefont {Yi}, \citenamefont {Chu},\ and\ \citenamefont
  {Mertz}(2006)}]{Yi2006}%
  \BibitemOpen
  \bibfield  {author} {\bibinfo {author} {\bibfnamefont {R.}~\bibnamefont
  {Yi}}, \bibinfo {author} {\bibfnamefont {K.~K.}\ \bibnamefont {Chu}}, \ and\
  \bibinfo {author} {\bibfnamefont {J.}~\bibnamefont {Mertz}},\ }\bibfield
  {title} {\enquote {\bibinfo {title} {Graded-field microscopy with white
  light},}\ }\href@noop {} {\bibfield  {journal} {\bibinfo  {journal} {Optics
  express}\ }\textbf {\bibinfo {volume} {14}},\ \bibinfo {pages} {5191--5200}
  (\bibinfo {year} {2006})}\BibitemShut {NoStop}%
\bibitem [{\citenamefont {Nomerotski}(2019)}]{Nomerotski2019}%
  \BibitemOpen
  \bibfield  {author} {\bibinfo {author} {\bibfnamefont {A.}~\bibnamefont
  {Nomerotski}},\ }\bibfield  {title} {\enquote {\bibinfo {title} {Imaging and
  time stamping of photons with nanosecond resolution in timepix based optical
  cameras},}\ }\href@noop {} {\bibfield  {journal} {\bibinfo  {journal}
  {Nuclear Instruments and Methods in Physics Research Section A: Accelerators,
  Spectrometers, Detectors and Associated Equipment}\ }\textbf {\bibinfo
  {volume} {937}},\ \bibinfo {pages} {26--30} (\bibinfo {year}
  {2019})}\BibitemShut {NoStop}%
\bibitem [{\citenamefont {Ianzano}\ \emph {et~al.}(2020)\citenamefont
  {Ianzano}, \citenamefont {Svihra}, \citenamefont {Flament}, \citenamefont
  {Hardy}, \citenamefont {Cui}, \citenamefont {Nomerotski},\ and\ \citenamefont
  {Figueroa}}]{Ianzano2020}%
  \BibitemOpen
  \bibfield  {author} {\bibinfo {author} {\bibfnamefont {C.}~\bibnamefont
  {Ianzano}}, \bibinfo {author} {\bibfnamefont {P.}~\bibnamefont {Svihra}},
  \bibinfo {author} {\bibfnamefont {M.}~\bibnamefont {Flament}}, \bibinfo
  {author} {\bibfnamefont {A.}~\bibnamefont {Hardy}}, \bibinfo {author}
  {\bibfnamefont {G.}~\bibnamefont {Cui}}, \bibinfo {author} {\bibfnamefont
  {A.}~\bibnamefont {Nomerotski}}, \ and\ \bibinfo {author} {\bibfnamefont
  {E.}~\bibnamefont {Figueroa}},\ }\bibfield  {title} {\enquote {\bibinfo
  {title} {Fast camera spatial characterization of photonic polarization
  entanglement},}\ }\href@noop {} {\bibfield  {journal} {\bibinfo  {journal}
  {Scientific reports}\ }\textbf {\bibinfo {volume} {10}},\ \bibinfo {pages}
  {1--11} (\bibinfo {year} {2020})}\BibitemShut {NoStop}%
\bibitem [{\citenamefont {Zhang}\ \emph {et~al.}(2020)\citenamefont {Zhang},
  \citenamefont {England}, \citenamefont {Nomerotski}, \citenamefont {Svihra},
  \citenamefont {Ferrante}, \citenamefont {Hockett},\ and\ \citenamefont
  {Sussman}}]{Zhang2020}%
  \BibitemOpen
  \bibfield  {author} {\bibinfo {author} {\bibfnamefont {Y.}~\bibnamefont
  {Zhang}}, \bibinfo {author} {\bibfnamefont {D.}~\bibnamefont {England}},
  \bibinfo {author} {\bibfnamefont {A.}~\bibnamefont {Nomerotski}}, \bibinfo
  {author} {\bibfnamefont {P.}~\bibnamefont {Svihra}}, \bibinfo {author}
  {\bibfnamefont {S.}~\bibnamefont {Ferrante}}, \bibinfo {author}
  {\bibfnamefont {P.}~\bibnamefont {Hockett}}, \ and\ \bibinfo {author}
  {\bibfnamefont {B.}~\bibnamefont {Sussman}},\ }\bibfield  {title} {\enquote
  {\bibinfo {title} {Multidimensional quantum-enhanced target detection via
  spectrotemporal-correlation measurements},}\ }\href {\doibase
  10.1103/PhysRevA.101.053808} {\bibfield  {journal} {\bibinfo  {journal}
  {Phys. Rev. A}\ }\textbf {\bibinfo {volume} {101}},\ \bibinfo {pages}
  {053808} (\bibinfo {year} {2020})}\BibitemShut {NoStop}%
\bibitem [{\citenamefont {Vidyapin}\ \emph {et~al.}(2022)\citenamefont
  {Vidyapin}, \citenamefont {Zhang}, \citenamefont {England},\ and\
  \citenamefont {Sussman}}]{vidyapin_characterisation_2022}%
  \BibitemOpen
  \bibfield  {author} {\bibinfo {author} {\bibfnamefont {V.}~\bibnamefont
  {Vidyapin}}, \bibinfo {author} {\bibfnamefont {Y.}~\bibnamefont {Zhang}},
  \bibinfo {author} {\bibfnamefont {D.}~\bibnamefont {England}}, \ and\
  \bibinfo {author} {\bibfnamefont {B.}~\bibnamefont {Sussman}},\ }\href
  {\doibase 10.48550/arXiv.2211.13788} {\enquote {\bibinfo {title}
  {Characterisation of a single photon event camera for quantum imaging},}\ }
  (\bibinfo {year} {2022}),\ \bibinfo {note} {arXiv:2211.13788 [physics,
  physics:quant-ph]}\BibitemShut {NoStop}%
\bibitem [{\citenamefont {England}, \citenamefont {Balaji},\ and\ \citenamefont
  {Sussman}(2019)}]{England2019}%
  \BibitemOpen
  \bibfield  {author} {\bibinfo {author} {\bibfnamefont {D.~G.}\ \bibnamefont
  {England}}, \bibinfo {author} {\bibfnamefont {B.}~\bibnamefont {Balaji}}, \
  and\ \bibinfo {author} {\bibfnamefont {B.~J.}\ \bibnamefont {Sussman}},\
  }\bibfield  {title} {\enquote {\bibinfo {title} {Quantum-enhanced standoff
  detection using correlated photon pairs},}\ }\href@noop {} {\bibfield
  {journal} {\bibinfo  {journal} {Physical Review A}\ }\textbf {\bibinfo
  {volume} {99}},\ \bibinfo {pages} {023828} (\bibinfo {year}
  {2019})}\BibitemShut {NoStop}%
\bibitem [{\citenamefont {Ndagano}\ \emph {et~al.}(2022)\citenamefont
  {Ndagano}, \citenamefont {Defienne}, \citenamefont {Branford}, \citenamefont
  {Shah}, \citenamefont {Lyons}, \citenamefont {Westerberg}, \citenamefont
  {Gauger},\ and\ \citenamefont {Faccio}}]{ndagano_quantum_2022}%
  \BibitemOpen
  \bibfield  {author} {\bibinfo {author} {\bibfnamefont {B.}~\bibnamefont
  {Ndagano}}, \bibinfo {author} {\bibfnamefont {H.}~\bibnamefont {Defienne}},
  \bibinfo {author} {\bibfnamefont {D.}~\bibnamefont {Branford}}, \bibinfo
  {author} {\bibfnamefont {Y.~D.}\ \bibnamefont {Shah}}, \bibinfo {author}
  {\bibfnamefont {A.}~\bibnamefont {Lyons}}, \bibinfo {author} {\bibfnamefont
  {N.}~\bibnamefont {Westerberg}}, \bibinfo {author} {\bibfnamefont {E.~M.}\
  \bibnamefont {Gauger}}, \ and\ \bibinfo {author} {\bibfnamefont
  {D.}~\bibnamefont {Faccio}},\ }\bibfield  {title} {\enquote {\bibinfo {title}
  {Quantum microscopy based on {H}ong–{O}u–{M}andel interference},}\ }\href
  {\doibase 10.1038/s41566-022-00980-6} {\bibfield  {journal} {\bibinfo
  {journal} {Nature Photonics}\ }\textbf {\bibinfo {volume} {16}},\ \bibinfo
  {pages} {384--389} (\bibinfo {year} {2022})},\ \bibinfo {note} {number: 5
  Publisher: Nature Publishing Group}\BibitemShut {NoStop}%
\bibitem [{\citenamefont {Torre}\ \emph {et~al.}(2022)\citenamefont {Torre},
  \citenamefont {McMillan}, \citenamefont {Monroy-Ruz},\ and\ \citenamefont
  {Matthews}}]{torre_sub-mum_2022}%
  \BibitemOpen
  \bibfield  {author} {\bibinfo {author} {\bibfnamefont {C.}~\bibnamefont
  {Torre}}, \bibinfo {author} {\bibfnamefont {A.}~\bibnamefont {McMillan}},
  \bibinfo {author} {\bibfnamefont {J.}~\bibnamefont {Monroy-Ruz}}, \ and\
  \bibinfo {author} {\bibfnamefont {J.~C.~F.}\ \bibnamefont {Matthews}},\
  }\href {\doibase 10.48550/arXiv.2212.02990} {\enquote {\bibinfo {title}
  {Sub-\{{\textbackslash}mu\}m axial precision depth imaging with entangled
  two-colour {Hong}-{Ou}-{Mandel} microscopy},}\ } (\bibinfo {year} {2022}),\
  \bibinfo {note} {arXiv:2212.02990 [physics, physics:quant-ph]}\BibitemShut
  {NoStop}%
\bibitem [{\citenamefont {Morimoto}\ \emph {et~al.}(2020)\citenamefont
  {Morimoto}, \citenamefont {Ardelean}, \citenamefont {Wu}, \citenamefont
  {Ulku}, \citenamefont {Antolovic}, \citenamefont {Bruschini},\ and\
  \citenamefont {Charbon}}]{Morimoto2020}%
  \BibitemOpen
  \bibfield  {author} {\bibinfo {author} {\bibfnamefont {K.}~\bibnamefont
  {Morimoto}}, \bibinfo {author} {\bibfnamefont {A.}~\bibnamefont {Ardelean}},
  \bibinfo {author} {\bibfnamefont {M.-L.}\ \bibnamefont {Wu}}, \bibinfo
  {author} {\bibfnamefont {A.~C.}\ \bibnamefont {Ulku}}, \bibinfo {author}
  {\bibfnamefont {I.~M.}\ \bibnamefont {Antolovic}}, \bibinfo {author}
  {\bibfnamefont {C.}~\bibnamefont {Bruschini}}, \ and\ \bibinfo {author}
  {\bibfnamefont {E.}~\bibnamefont {Charbon}},\ }\bibfield  {title} {\enquote
  {\bibinfo {title} {Megapixel time-gated spad image sensor for 2d and 3d
  imaging applications},}\ }\href {\doibase 10.1364/OPTICA.386574} {\bibfield
  {journal} {\bibinfo  {journal} {Optica}\ }\textbf {\bibinfo {volume} {7}},\
  \bibinfo {pages} {346--354} (\bibinfo {year} {2020})}\BibitemShut {NoStop}%
\bibitem [{Can()}]{Canon}%
  \BibitemOpen
  \href {https://global.canon/en/news/2021/20211215.html} {\enquote {\bibinfo
  {title} {https://global.canon/en/news/2021/20211215.html},}\ }\BibitemShut
  {NoStop}%
\bibitem [{\citenamefont {Mertz}(2019)}]{Mertz2019}%
  \BibitemOpen
  \bibfield  {author} {\bibinfo {author} {\bibfnamefont {J.}~\bibnamefont
  {Mertz}},\ }\href@noop {} {\emph {\bibinfo {title} {Introduction to optical
  microscopy}}}\ (\bibinfo  {publisher} {Cambridge University Press},\ \bibinfo
  {year} {2019})\BibitemShut {NoStop}%
\bibitem [{\citenamefont {Lowenthal}\ and\ \citenamefont
  {Belvaux}(1967)}]{Lowenthal1967}%
  \BibitemOpen
  \bibfield  {author} {\bibinfo {author} {\bibfnamefont {S.}~\bibnamefont
  {Lowenthal}}\ and\ \bibinfo {author} {\bibfnamefont {Y.}~\bibnamefont
  {Belvaux}},\ }\bibfield  {title} {\enquote {\bibinfo {title} {Observation of
  phase objects by optically processed hilbert transform},}\ }\href@noop {}
  {\bibfield  {journal} {\bibinfo  {journal} {Applied Physics Letters}\
  }\textbf {\bibinfo {volume} {11}},\ \bibinfo {pages} {49--51} (\bibinfo
  {year} {1967})}\BibitemShut {NoStop}%
\bibitem [{\citenamefont {Couteau}(2018)}]{Couteau2018}%
  \BibitemOpen
  \bibfield  {author} {\bibinfo {author} {\bibfnamefont {C.}~\bibnamefont
  {Couteau}},\ }\bibfield  {title} {\enquote {\bibinfo {title} {Spontaneous
  parametric down-conversion},}\ }\href@noop {} {\bibfield  {journal} {\bibinfo
   {journal} {Contemporary Physics}\ }\textbf {\bibinfo {volume} {59}},\
  \bibinfo {pages} {291--304} (\bibinfo {year} {2018})}\BibitemShut {NoStop}%
\bibitem [{\citenamefont {Dudak}(2020)}]{Dudak2020}%
  \BibitemOpen
  \bibfield  {author} {\bibinfo {author} {\bibfnamefont {J.}~\bibnamefont
  {Dudak}},\ }\bibfield  {title} {\enquote {\bibinfo {title} {High-resolution
  x-ray imaging applications of hybrid-pixel photon counting detectors
  timepix},}\ }\href@noop {} {\bibfield  {journal} {\bibinfo  {journal}
  {Radiation Measurements}\ }\textbf {\bibinfo {volume} {137}},\ \bibinfo
  {pages} {106409} (\bibinfo {year} {2020})}\BibitemShut {NoStop}%
\end{thebibliography}%

\end{document}